\documentclass[11pt,5p,twocolumn,final]{elsarticle}
\journal{Journal of Molecular Spectroscopy}

\usepackage{multirow}
\usepackage{textcomp}
\usepackage{epstopdf}
\newcommand*\rot{\rotatebox{90}}

\begin{document}

\begin{frontmatter}

\title{Radiative branching ratios for excited states of  $^{174}$YbF: application to laser cooling}
\author[ICL]{I.J. Smallman}
\author[ASU]{F. Wang}
\author[ASU]{T.C. Steimle}
\author[ICL]{M.R. Tarbutt}
\author[ICL]{E.A. Hinds\corref{cor}}
\ead{ed.hinds@imperial.ac.uk}
\cortext[cor]{Corresponding author.}
\address[ICL]{Centre for Cold Matter, Blackett Laboratory, Imperial College London, \\Prince Consort Road, London SW7 2AZ, UK.}
\address[ASU]{Department of Chemistry and Biochemistry, Arizona State University, \\Tempe, Arizona 85287-1604, USA.}

\begin{abstract}
We excite YbF molecules to low-lying vibrational levels of the $A^2\Pi_{1/2}$ state, and of the nearby perturber state sometimes called [18.6]0.5.  By dispersing the fluorescence, we measure branching ratios for the radiative decay to vibrational levels of the $X^2\Sigma^+$ state. These ratios help to determine the optimum laser cooling scheme for the molecule. Our results establish the practicality of a scheme that was previously suggested \mbox{[Tarbutt \textit{et al.}, New J. Phys. 15 (2013) 053034]} and show how the scheme can be modified to provide a stronger radiative cooling force.
\end{abstract}

\begin{keyword}
radiative branching ratios \sep ytterbium fluoride (YbF)\sep laser cooling molecules
\end{keyword}

\end{frontmatter}

\section{Introduction}
\label{sec:Intro}

The scattering of laser light is now well established as the standard way to cool atoms to micro-Kelvin temperatures. By contrast, the laser cooling of molecules is only at an early stage of development, the main hurdle being the need for spontaneous scattering of many photons. In molecules, an electronically excited state with vibrational quantum number $v'$ normally decays to many different vibrational levels $v''$ of the ground electronic state, with branching ratios $b_{v',v''}$. Typically these factors permit only a few optical excitations before the vibrational state changes and the molecule is no longer excited by the light.  In some molecules, however, the optically active electron is scarcely involved in the bonding and the matrix of branching ratios is nearly diagonal  \cite{DiRosa2004}, i.e. $b_{v,v}\simeq1$. Moreover, the small chance of changing the vibrational state greatly favours $\Delta v=1$, with $\Delta v=2$ being quadratically small etc.  For these molecules it is possible to return lost population back into the laser cooling cycle using only one or two additional re-pumping lasers. Following an initial demonstration of the radiative force acting on SrF molecules \cite{Shuman2009}, transverse laser cooling was applied to beams of SrF \cite{Shuman2010} and YO \cite{Hummon2013} molecules. Radiation pressure has also been used to slow down beams of SrF \cite{Barry2012}, and recently a beam of CaF has been both slowed and longitudinally cooled \cite{Zhelyazkova2013}.

\begin{figure}[t]
        \centering
        {
        	\includegraphics{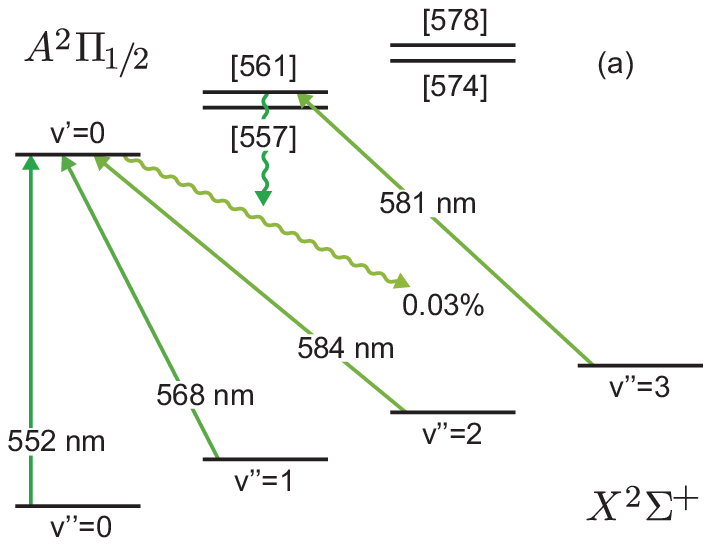}
             \label{fig:EnergyLevelDiagramA}
        }
        \vspace{0.4cm}
     {
        	\includegraphics{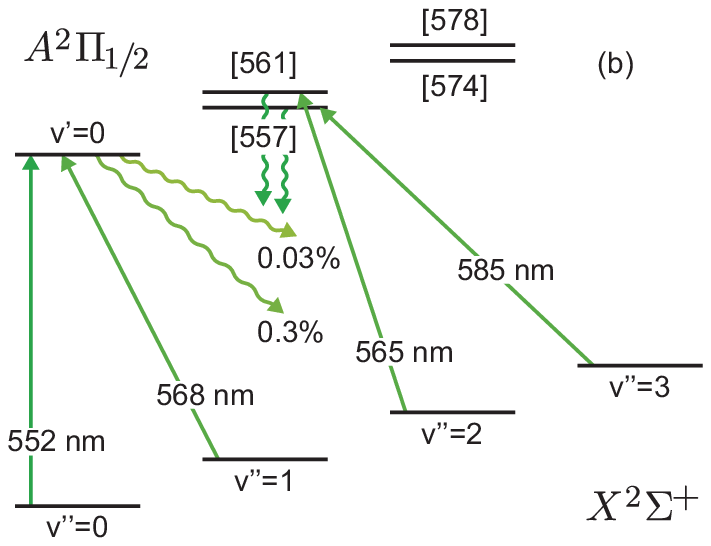}
             \label{fig:EnergyLevelDiagramB}
        }
        \caption{ $^{174}$YbF energy levels relevant for laser cooling. The numbers in  square brackets indicate energy above \mbox{$X^2\Sigma^{+}(v''=0)$} in terahertz. (a) Laser cooling scheme for YbF proposed in ref. \cite{Tarbutt2013}. (b) Improved scheme resulting from the present measurements.}
        \label{fig:EnergyLevelDiagram}
\end{figure}

The molecule $^{174}$YbF is of particular interest for elementary particle physics and cosmology as it provides a strong constraint on the permanent electric dipole moment of the electron (eEDM) \cite{Hudson2011,Kara2012}. Because the electronic structure of this molecule resembles that of SrF and CaF,  it is also a candidate for laser cooling. Indeed, a recent measurement of branching ratios for the decay from \mbox{$A^2\Pi_{1/2}(N'=0,v'=0)$} in $^{174}$YbF \cite{Zhuang2011} showed that it could be practical to laser-cool this molecule.  This led to a proposal for making a laser-cooled fountain of $^{174}$YbF molecules, which would permit long coherent observation times, thereby improving the eEDM measurement by a factor of a thousand \cite{Tarbutt2013}. The scheme proposed in  \cite{Tarbutt2013} and illustrated in figure \ref{fig:EnergyLevelDiagram}(a) involves the electric dipole transition between the even-parity upper level \mbox{$A^2\Pi_{1/2}(N'=0, v'=0)$} and the odd-parity lower level \mbox{$X^2\Sigma^+(N''=1, v'' = 0)$}.  Molecules in the excited state have a 6.6\% probability of decay to \mbox{$v''=1$} and a 0.3\% branch to $v''=2$. Since several thousand photons must be scattered to cool the molecules effectively, repump lasers return $v''=1$ and $v''=2$ molecules to the cooling cycle through the A-state. Within this four-level system, there are nine X-state sublevels (3 projections of $N''=1$ in each of $v''=0,1,2$) for every A-state sublevel.  Reference \cite{Tarbutt2013} shows that strong laser cooling populates all the sublevels equally, leading in this case to a maximum scattering rate of approximately $\Gamma/10$, where $\Gamma$ is the inverse of the A-state radiative lifetime. 

Figure \ref{fig:EnergyLevelDiagram}(b) shows a possible improvement to the scheme. Here, only the $v''=0$ and $v''=1$ states are coherently coupled to the upper level $v'=0$, so the scattering rate increases to $\Gamma/7$ and the cooling is correspondingly stronger. Molecules that fall into  $v''=2$ are now repumped through the first vibrationally excited  A-state and relax incoherently back into the cooling cycle. However, this vibrationally excited state in YbF is strongly mixed with a perturbing state, sometimes called [18.6]0.5 \cite{Dunfield1995,Sauer1999}, resulting in two states, which we label by their energy in THz relative to X$(v''=0)$ as  $[557]$ and $[561]$. While the X and A states of YbF come from the Yb$^{+}$ $4f^{14}6s$ and $4f^{14}6p$ configurations respectively, the perturber is thought to come from the Yb$^{+}$ $4f^{13}6s^{2}$ configuration \cite{Sauer1999}. Because of this change to the nature of the bond, it seemed likely that the branching ratios for both these 
strongly-mixed states would be unfavourable for laser cooling, and therefore the scheme of figure \ref{fig:EnergyLevelDiagram}(b) was not proposed in \cite{Tarbutt2013}. In proposing a weak repumper for $v''=3$ molecules via $[561]$ reference \cite{Tarbutt2013} took a chance that the branching ratio for this state would not be too unfavourable.  

In this paper, we describe measurements of the branching ratios for decay from both $[557]$ and $[561]$, and also from the next vibrationally excited doublet, $[574]$ and $[578]$. We show that, unexpectedly,  $[557]$ and $[561]$ are very well suited for re-pumping the vibrationally excited molecules and, on the basis of our results, we conclude that laser cooling will work well with the scheme shown in figure \ref{fig:EnergyLevelDiagram}(b).

\begin{figure}[h]
    \centering
    \includegraphics{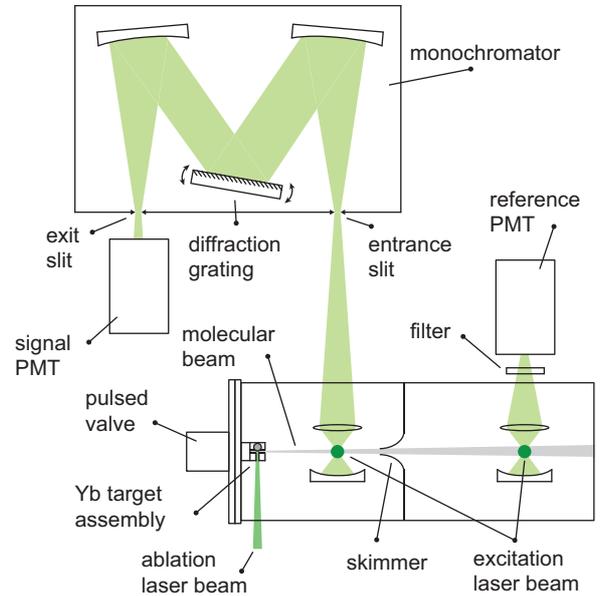}
    \caption{Schematic diagram of apparatus. A supersonic jet beam of YbF molecules is illuminated by a laser beam between the source and the skimmer. The resulting fluorescence is imaged into a scanning monochromator, which disperses the spectrum and measures the relative strengths of the decay branches to the various vibrational levels of the electronic ground state.  A second excitation beam downstream from the skimmer provides a reference signal to compensate for drifts of the molecular beam intensity.}
  \label{fig:Apparatus}
\end{figure}

\section{Experiment}
\label{sec:Exp}

\begin{figure*}[t]
	\centering
        \includegraphics{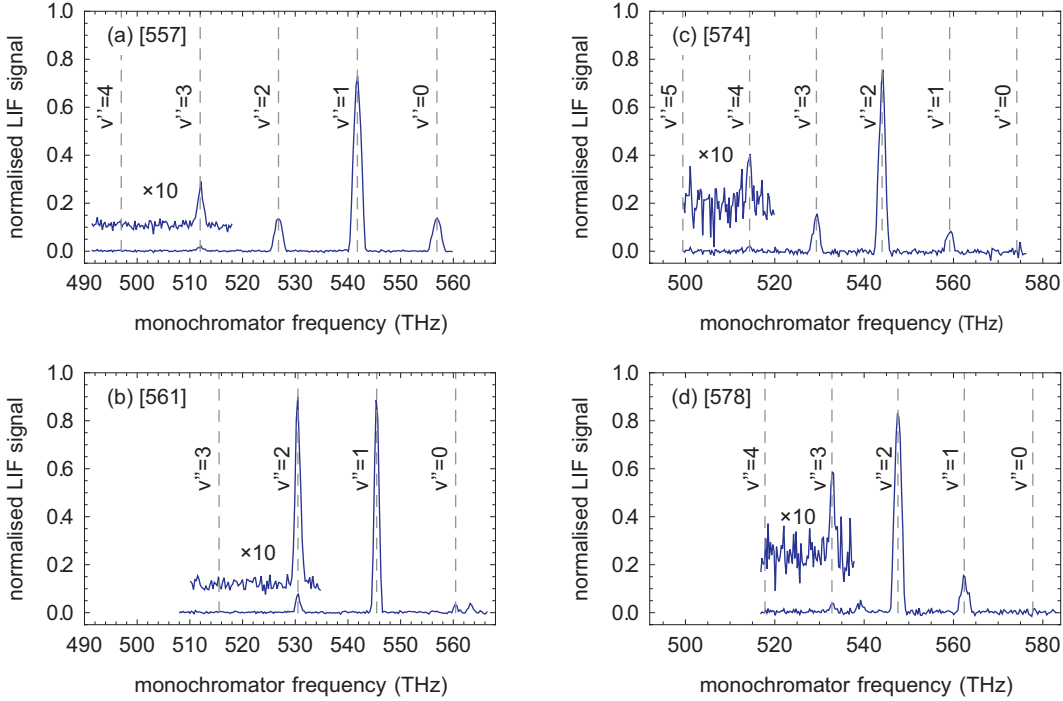}
         \caption{Dispersed laser-induced fluorescence spectra measured by a scanning monochromator. The four excited states investigated are (a) $[557]$, (b) $[561]$, (c) $[574]$, and (d) $[578]$. The fluorescence signals are normalised such that the peak heights indicate the measured branching ratios. The smaller peaks are made more evident by re-plotting them with an offset and with a tenfold magnification.}
    \label{fig:DispersedFluorescenceSpectra}
\end{figure*}

The experiments,  performed at Arizona State University, use the molecular beam and spectrometer illustrated schematically in figure \ref{fig:Apparatus}. Following the method of reference \cite{Tarbutt2002}, the YbF molecules are produced in a pulsed beam with a temperature of a few Kelvin. A 27\,bar mixture of 98\% Ar and 2\% SF$_{6}$ is fed to a solenoid valve. This opens every 50\,ms to release 100--200\,\textmu s-long gas pulses into the vacuum chamber, where the pressure is kept below 10$^{-4}$\,mbar. Immediately outside the valve, a rod of Yb metal is held next to a narrow channel through which the gas mixture flows. Just as each gas pulse reaches its highest density in the channel, a pulse of light is focussed onto the rod  in order to ablate the Yb. The light, from a frequency-doubled, Q-switched Nd:YAG laser, has a pulse energy of  $\sim$5\,mJ/pulse and a frequency of 564\,THz. The ablated Yb reacts with the SF$_{6}$ to produce YbF molecules that are entrained in the gas and form a beam. In order to maintain a high yield of molecules, the surface of the Yb is continuously renewed by rotating and translating the rod. Approximately 30\,cm downstream from the source, the beam is collimated by the 2\,mm aperture of a skimmer and passes into a second vacuum chamber, where the pressure is below 10$^{-7}$\,mbar.

\begin{table*}[t]
	\centering
	\caption{Measured branching ratios (\%) for radiative decay from the  $[557]$, $[561]$, $[574]$ and  $[578]$ states of $^{174}$YbF to the state $X^2\Sigma^{+}(v'')$.}
	\begin{tabular}{ c  c | c c c c c c |}
		\cline{3-8}
		& & \multicolumn{6}{c|}{\mbox{$X^2\Sigma^{+}$}} \\
		\cline{3-8}
		& & $v''=0$ & $v''=1$ & $v''=2$ & $v''=3$ & $v''=4$ & $v''=5$ \\ 
		\hline 
		\multicolumn{1}{|c|}{\multirow{4}{*}{\rot{upper state\,\,\,}}} 
			& $[557]$ & $13.2\pm0.5$ & $70.7\pm0.6$ & $13.9\pm0.2$ & $1.9\pm0.2$ & $<0.2$ &\\ 
		\multicolumn{1}{|c|}{} & $[561]$ & $2.8\pm0.3$ & $89\pm1$ & $7.8\pm0.2$ & $<0.2$ & & \\
\multicolumn{1}{ |c| }{}&{$[574]$} & $<1$ & $8.1\pm0.6$ & $75\pm2$ & $15\pm2$ & $1.9\pm0.8$ & $<1$ \\
\multicolumn{1}{ |c| }{}&{$[578]$} & $<1$ & $13.3\pm0.6$ & $82.5\pm0.7$ & $4.1\pm0.6$ & $<1$ & \\
		\hline
	\end{tabular}
	\label{tab:FCfactors}
\end{table*}

We use a cw, single-mode dye laser to excite molecules from the ground state  \mbox{$X^2\Sigma^+(v'' = 0)$}. The  laser beam crosses the molecular beam in the free jet expansion region between the valve and the skimmer, as shown in figure \ref{fig:Apparatus}.  The fluorescence from these excited molecules is focussed into a 0.66\,m scanning monochromator, adjusted to have a spectral resolution of 1.2\,nm. We obtain the  dispersed spectra shown in figure \ref{fig:DispersedFluorescenceSpectra} by scanning the grating and recording many beam shots. Looking at figures \ref{fig:DispersedFluorescenceSpectra}(a) and (b), we see that $[557]$ and $[561]$ have their strongest decay to $v''=1$, and that there are also weaker branches to $v''=0,\,2$ and $3$. The scales are chosen so that the height of each peak indicates the branching ratio for that transition. We note that the small peak at 564\,THz in figure \ref{fig:DispersedFluorescenceSpectra}(b) is due to scattered Nd:YAG light that escaped from the laser cavity while it was nominally closed. Three corrections have gone into figures \ref{fig:DispersedFluorescenceSpectra}(a) and (b). (i) The spectrometer is calibrated using a tungsten filament having a known power spectrum, which allows us to rescale the fluorescence according to its wavelength to obtain the correct relative intensities of the YbF lines. (ii) When we block the Nd:YAG laser to measure a background spectrum without molecules, we find a small small $v''=0$ peak due to laser light that is scattered by the apparatus, rather than the molecules. This background spectrum has been subtracted. (iii) The cw laser beam is split into two so that a second beam can excite the molecules  at a point downstream from the skimmer, as shown in figure \ref{fig:Apparatus}. Here, the integrated fluorescence from each beam shot is detected by a photomultiplier, allowing us to monitor the drift  of the molecular beam intensity and hence to correct for it in the fluorescence spectrum.  A filter in front of the reference PMT blocks light at the laser frequency, thereby avoiding the potential problem of light scattered by the optics.
 
The fluorescence spectra from $[574]$ and $[578]$ are shown in figures \ref{fig:DispersedFluorescenceSpectra}(c) and \ref{fig:DispersedFluorescenceSpectra}(d). In both cases the favoured decay is to $v''=2$, accompanied by weaker branches to $v''=1$ and $v''=3$. The branches to $v''=0$ and $v''\ge4$ are very small, if present at all. Through careful adjustment of the source we ensure that no  Nd:YAG light is scattered into the monochromator, which is important because 564\,THz is almost identical to the frequency of the $v''=1$ peaks in these two cases. We do not normalise these spectra for drifts in molecular beam intensity, but as before, we do calibrate the scans with a tungsten filament and subtract the background spectrum.

The spectra in figure \ref{fig:DispersedFluorescenceSpectra} are averages over at least three scans, with each scan having a point every 0.3\,nm ($\sim$0.3\,THz) and 10 beam shots per point. The spectra in (a) and (b) are less noisy than those in (c) and (d) because the probabilities for exciting  into $[557]$ or $[561]$ from \mbox{$v''=0$} are much greater than those for $[574]$ and $[578]$. The measured branching ratios are listed in table \ref{tab:FCfactors}. Four lines are visible in the fluorescence from $[557]$, and branching ratios for these are derived from each of three spectra. The numbers presented in the table are the mean values, with error bars being the observed standard deviation of the mean. These uncertainties are consistent with shot noise due to photon counting statistics. No line can be seen at $v''=4$ (figure \ref{fig:DispersedFluorescenceSpectra}(a)), but by averaging this region over eight scans we are able to place an upper limit of 0.2\% on the strength of this branch. The branching ratios decrease so rapidly with increasing $v''$ that this upper limit should also apply to the sum over \mbox{$v''\ge4$}. For the decay from $[561]$, weak  branches are seen in figure \ref{fig:DispersedFluorescenceSpectra}(b) to \mbox{$v''=0$} and \mbox{$v''=2$}. After averaging the \mbox{$v''=3$} region over 12 scans, we similarly assign an upper limit of 0.2\% both to the \mbox{$v''=3$} branch and to the sum over \mbox{$v''\ge3$}. In the fluorescence from the $[574]$ and $[578]$ states, we see branches to  \mbox{$v''=1-4$} and \mbox{$v''=1-3$} respectively. Although the decays to  \mbox{$v''=0$} are at the wavelength of the excitation laser, we are able to control the scattered laser light and to subtract the small residual background well enough to place an upper limit of 1\% on these branches. We also find no evidence for branching to \mbox{$v''\ge5$} and \mbox{$v''\ge4$}, so these too are assigned an upper limit of 1\%. 

\section{Discussion}

The motivation for our measurement of these branching ratios was to address two questions about the YbF laser cooling scheme proposed in \cite{Tarbutt2013} and shown in figure \ref{fig:EnergyLevelDiagram}(a). The first concerns repumping $v''=2$ through a different upper level, as in figure \ref{fig:EnergyLevelDiagram}(b). This would give higher maximum cooling power, and hence a higher capture velocity for stopping the molecules in optical molasses, but we needed to determine whether it would also compromise the cooling by pumping molecules more rapidly into \mbox{$v''=3$}. In the scheme of figure \ref{fig:EnergyLevelDiagram}(a), the probability of decay to \mbox{$v''=3$} is \mbox{$b_{0,3}=3 \times 10^{-4}$} for every photon scattered. The scheme shown in figure \ref{fig:EnergyLevelDiagram}(b), has an additional probability of leaking molecules into \mbox{$v''=3$} given approximately by \mbox{$b_{0,2}\times b_{[557],3}$} or by \mbox{$b_{0,2}\times b_{[561],3}$}, depending on which state is used for the repumping. With \mbox{$b_{0,2}=3 \times 10^{-3}$}  \cite{Zhuang2011}, and the values listed in table \ref{tab:FCfactors}, we now know that these additional probabilities for populating \mbox{$v''=3$} are approximately \mbox{$6\times 10^{-5}$} and \mbox{$<6\times 10^{-6}$} respectively.  Thus they are both small enough for scheme (b) to work well, with $[561]$ being the better choice because it returns the molecules more readily to the cooling cycle. We note that decay from $[561]$ to $[557]$ would take molecules out of the cooling cycle by leaving them with the wrong parity, however, the $\omega^3$ factor alone suppresses this branch by a factor of \mbox{$4\times 10^{-7}$}, so we do not expect it to be significant.
 
The second question is how to re-pump the \mbox{$v''=3$} molecules. In \cite{Tarbutt2013} it was assumed that this could be done a few times through $[557]$ or $[561]$  without losing too many  molecules to higher vibrational states. The measurements presented here show this to be true. Since we propose to use $[561]$ for re-pumping \mbox{$v''=2$}, it is best to use one of the other states to re-pump \mbox{$v''=3$}, so that coherent coupling between \mbox{$v''=2$} and \mbox{$v''=3$} is avoided.  Looking at the other three states in table \ref{tab:FCfactors}, the next most favourable one for re-pumping \mbox{$v''=3$} is $[557]$, which has less than $0.2\%$ probability of loss to \mbox{$v''>3$}. In that case, the main loss to \mbox{$v''=4$} will be the direct decay from \mbox{$v'=0$}, for which we estimate \mbox{$b_{0,4}\simeq 1\times10^{-5}$}, based on the trend of the branching ratios measured in \cite{Zhuang2011}. 

In conclusion, we have measured previously unknown radiative decay branching ratios in YbF in order to learn more about the possibility of laser cooling this molecule. Our results show that \mbox{$v''=3$} can indeed be repumped via $[557]$ or $[561]$ without undue loss of molecules to higher vibrational states, which was the hope in reference \cite{Tarbutt2013}. We have also shown that the \mbox{$v''=2$} state can be repumped via $[561]$ instead of $v'=0$, thereby permitting a 42\% stronger scattering force, with correspondingly higher stopping power.

\section*{Acknowledgements}

The authors would like to thank Anh Le for her assistance with the experimental measurements, as well as Jony Hudson and Ben Sauer for many useful discussions. The work at Arizona State University was supported by a grant (CHE-1265885) from the Chemistry Division of the National Science Foundation. Work at Imperial College London was supported by the Royal Society, EPSRC and ERC. 

\bibliographystyle{elsarticle-num}
\bibliography{References}

\end{document}